\documentclass[fleqn]{article}
\usepackage{longtable}
\usepackage{graphicx}

\begin {document}
\begin{flushleft}
{\LARGE
{\bf Comment on ``Atomic structure calculations for F-like tungsten" by S. Aggarwal  [Chin. Phys. B  23 (2014) 093203]}
}\\

\vspace{1.5 cm}

{\bf {Kanti  M.  ~Aggarwal}}\\ 

\vspace*{1.0cm}

Astrophysics Research Centre, School of Mathematics and Physics, Queen's University Belfast, Belfast BT7 1NN, Northern Ireland, UK\\ 
\vspace*{0.5 cm} 

e-mail: K.Aggarwal@qub.ac.uk \\

\vspace*{0.20cm}

Received  2 November 2015.  Accepted for publication xx Month 2015 \\

\vspace*{1.5cm}


\vspace*{1.0 cm}

\hrule

\vspace{0.5 cm}

\end{flushleft}

\clearpage


\begin{abstract}

Recently, S. Aggarwal [Chin. Phys. B  23 (2014) 093203] reported energy levels, radiative rates and lifetimes for the lowest 60 levels belonging to the 2s$^2$2p$^5$, 2s2p$^6$ and  2s$^2$2p$^4$3$\ell$ configurations of F-like tungsten. There is no discrepancy for his calculated energies for the levels and the radiative rates for the limited number of E1  transitions, but the reported results for lifetimes are highly inaccurate. According to our calculations, errors in his reported lifetimes are up to 6 orders of magnitude for several levels. Here we report the correct lifetimes for future comparisons and applications, and also explain the reason for discrepancies.

\end{abstract}

\clearpage

\section{Introduction}

Atomic data, including energy levels and  oscillator strengths or radiative decay rates,  are required for many ions in order to model plasmas and/or for diagnostics, such as to determine elemental density, temperature and chemical composition. Generally, lighter ions (with Z $\le$ 28) are important for astrophysical applications whereas heavier ones are more important for lasing and fusion plasmas. An important element of particular interest  for studies of fusion plasmas is tungsten (W), because apart from being an important constituent of the reactor walls, it radiates at several of its ionisation stages. Considering the requirement of the developing  ITER project, atomic data for its ions are in high demand. However, it is not any data, but accurate and complete set of data which are required for modelling and diagnostics$^{\cite{fst}}$. Recently, S. Aggarwal$^{\cite{sa}}$, henceforth to be referred to as AS,  has reported results for energy levels, oscillator strengths (f-values), radiative rates (A-values), line strengths (S-values),  and lifetimes ($\tau$) for F-like W~LXVI. For his calculations,  the modified version of the {\sc grasp} (general-purpose relativistic atomic structure package) code was adopted. This code is known as GRASP0 and is freely  available at the website {\tt http://web.am.qub.ac.uk/DARC/}. It is a fully relativistic code,  based on the $jj$ coupling scheme, and includes further relativistic corrections arising from the Breit interaction and QED (quantum electrodynamics) effects. 

SA reported energies and lifetimes  for the lowest 60 levels of the 2s$^2$2p$^5$, 2s2p$^6$ and  2s$^2$2p$^4$3$\ell$ configurations of W~LXVI, although he included {\em configuration interaction} (CI) among 359 levels. The additional 299 levels arise from the 2s$^2$2p$^4$4$\ell$, 2s$^2$2p$^4$5$\ell$  ($\ell \le$ 3), 2s2p$^5$3$\ell$, 2s2p$^5$4$\ell$, and 2p$^6$3$\ell$ configurations, i.e. 23 in total.  However, he listed A-values (and other related parameters) for only electric dipole (E1) transitions from the lowest two levels of the ground configuration (2s$^2$2p$^5$~$^2$P$^o_{3/2,1/2}$) to higher excited levels.  Such limited data are worthless because for modelling of plasmas a complete set of data, for all transitions, are required. Additionally, for the calculations of $\tau$ A-values are also required for the  electric quadrupole (E2), magnetic dipole (M1), and magnetic quadrupole (M2) transitions. Because of the limitations of his data, we notice that while his reported A-values are correct, the results for $\tau$ are not. Therefore, in this communication we report the correct values of lifetimes.


\section{Calculations}

For consistency, we have employed  the same version of the  {\sc grasp}  code as adopted by SA$^{\cite{sa}}$. Similarly,  for compatibility we use the same  option of {\em extended average level} (EAL), as adopted by him. To make comparisons, we have performed three sets of calculations, namely GRASP1: with 113 levels of 11 configurations, i.e. 2s$^2$2p$^5$, 2s2p$^6$,  2s$^2$2p$^4$3$\ell$, 2s2p$^5$3$\ell$, and 2p$^6$3$\ell$; GRASP2: which includes the same 23 configurations and 359 levels as by SA (see the list in section 1); and finally GRASP3: which includes further 142 levels of the 2s$^2$2p$^4$5g,  2s2p$^5$5$\ell$, 2p$^6$4$\ell$, and 2p$^6$5$\ell$ configurations. Therefore, GRASP3 calculations include 501 levels from 38 configurations in total. This exercise is performed mainly to (i) demonstrate that inclusion of CI additional to the basic 11 configurations of GRASP1 makes a nominal impact on the calculations of atomic parameters, including $\tau$, and (ii) levels of higher configurations intermix with the lowest 113. There are no  (major) differences  for energy levels, as may be noted from Table~1 in which we list our energies from GRASP3 and those of SA (designated as GRASP4 for convenience although he includes the same levels as in GRASP2).  However, results are listed for only the lowest 92 levels, because beyond those levels of the 2s$^2$2p$^4$4s configuration intermix (see Table~8 of Aggarwal and Keenan$^{\cite{ak}}$).  Similarly, there are no differences in  A-values between our calculations and those of SA. However, we note here that energies obtained by SA with the {\em Flexible Atomic Code} (FAC) of Gu$^{\cite{gu}}$ and listed in his Table~2 are {\em incorrect} for some of the levels, particularly the last 5, because these differ with his GRASP calculations by nearly 1 Ryd. We have performed similar calculations with FAC and do not observe any such discrepancies. Furthermore, it is clear from Table~1 that there are many levels missing from the tabulation by SA. It is not apparent from his paper why he excluded these levels, in spite of including in the calculations. We now focus  on the results for $\tau$ for which there are large discrepancies.

\section{Lifetimes}

The lifetime $\tau$ of a level $j$ is determined as  ${\tau}_j =1/{\sum_{i} A_{ji}}$, i.e. the summation is over {\em all}\, types of transitions from lower levels $i$ to higher $j$.  In Table~1, we list our calculated $\tau$ from  GRASP1, GRASP2  and GRASP3 calculations, along with those of SA (GRASP4) for a ready comparison.  However, results are listed for only the lowest 92 levels, because beyond those  intermixing of the levels of the 2s$^2$2p$^4$4s configuration affect  the calculations of $\tau$. However, our calculated $\tau$ for the last two levels of SA, namely (2s$^2$2p$^4$3d)~$^4$F$_{3/2}$ and $^2$F$_{5/2}$, not included in our Table~1, are 8.119$\times$10$^{-16}$ and 2.195$\times$10$^{-12}$~s, respectively, and correspond to levels 101 and 106 listed in Table~8 of Aggarwal and Keenan$^{\cite{ak}}$. For these two levels the respective $\tau$ by SA are 8.65$\times$10$^{-10}$ and 1.00$\times$10$^{-11}$~s, i.e. differing by up to 6 orders of magnitude.

It is clear from Table~1 that all three calculations, namely GRASP1, GRASP2 and GRASP3, determine values of $\tau$ within a satisfactory agreement, and the discrepancy, if any, is below 10\% -- see level 33 for example. Therefore, as stated above the effect of additional CI included in GRASP2 and GRASP3 calculations is nominal on the determination of atomic parameters. On the other hand, the $\tau$ results of SA are invariably  {\em higher},  by up to 6 orders of magnitude for several levels -- see for example 14, 41, 52, 56, and 59. Since SA included the same CI as in our GRASP2 calculations, his results for $\tau$ should have been comparable with ours. Unfortunately, this is not the case, and the discrepancy cannot be understood because of the limited listing of A-values by SA. In this short communication it is not possible for us to list A-values for all transitions, but in Table~2 we  list all those A-values which contribute to over 20\% to ${\sum_{i} A_{ji}}$. This will help us to understand the differences between ours and the $\tau$ of SA. We also note that for level 60 (2s$^2$2p$^4$3d~$^2$D$_{5/2}$), several transitions make  substantial contributions but individually below 20\% of the sum total. Therefore, for this level those transitions are listed which contribute to over 10\%. Finally, all A-values for a wide range of transitions are provided in a separate paper$^{\cite{ak}}$.

SA mentioned about the calculations for M2 transitions (only in the Abstract but not in the text of the paper), and nothing about  E2 and M1. Additionally, he  presented no  results for any other types except E1. However, it is  clear from Table~2 that for the determination of $\tau$ for level 2, the 1--2 M1 transition dominates. Since there is no discrepancy between the two sets of results for this level (see Table~1),  he {\em must} have included the contributions of M1 transitions. However,  he did not calculate A-values for E2 transitions, because for 7--9 levels, their A-values  clearly dominate, and hence the discrepancies between the GRASP2 and GRASP4 results for $\tau$. On the other hand, for level 20 the 1--20 M2 and 10--20 E1 transitions contribute almost equally. As there is a discrepancy of over an order of magnitude in $\tau$ for this level, it is clear that he made no calculations for M2 transitions. However, the 10--20 E1 A-value (9.891$\times$10$^{10}$ s$^{-1}$) should give $\tau$ = 1.01$\times$10$^{-11}$ s and not 9.36$\times$10$^{-10}$ s, as listed by him. This is because, his calculation is based on A-value (1.068$\times$10$^{9}$ s$^{-1}$) for the 8--20 E1 transition alone. Among the E1 transitions, the one which  dominates is 12--20 with A = 2.747$\times$10$^{10}$ s$^{-1}$, not listed in Table~2.  Therefore, apart from the omissions of E2 and M2 transitions by him, there are other reasons too.  We discuss this further.

For level 14, the 6--14 E1 A-value (1.476$\times$10$^{12}$ s$^{-1}$) contributes to over 76\%, but the $\tau$ of SA is perhaps based on 3--14 (A= 3.80$\times$10$^9$, s$^{-1}$) alone, and hence the discrepancy of four orders of magnitude. Similarly, for level 56 the A-value for the 38--56 E1 transition contributes to about half, but the calculation of SA is probably based on the A-value for the  8--56 E1 (A= 2.446$\times$10$^6$ s$^{-1}$) transition alone. Therefore, it is apparent that his calculations are totally random and hence highly unreliable. It may be worth noting here that W~LXVI is not the only ion for which we have found discrepancies, errors and anomalies in his (and the group members) reported work. Several of their calculations have been demonstrated as unreliable -- see for example W~XL$^{\cite{w40}}$, W~LVIII$^{\cite{w58}}$ and W~LXII$^{\cite{w62}}$.

\section{Conclusions}

In this work, we have calculated energy levels, radiative rates and lifetimes for the lowest 501 levels/transitions of F-like tungsten, i.e. W LXVI, by adopting the widely used {\sc grasp} code. However, results are presented here only for the lowest 92 levels to make comparisons with the recent work of  S. Aggarwal$^{\cite{sa}}$, who has  reported similar results. His results for energy levels and A-values, only  for limited levels/transitions,  are found to be correct, but the corresponding values of $\tau$ are in large errors, of up to 6 orders of magnitude, for several levels. This is because he has randomly selected transitions for the calculations, which are always not the most contributing ones. Therefore, for the benefit of users (mainly experimentalists) as well as for future workers, we have listed the correct values of $\tau$, and have also explained the reasons for discrepancies. 

Since S.~Aggarwal$^{\cite{sa}}$ reported  A-values only for E1 transitions from the ground levels, these are insufficient for applications, because a {\em complete} set of data for {\em all} transitions are required for any modelling application. Additionally, he did not consider all levels of the calculations as several are missing (see Table~1). Furthermore, there is scope for improvement in his work by including additional CI in the calculations. Therefore, an improved set of complete data for all transitions of W LXVI are reported in a separate paper$^{\cite{ak}}$.




\newpage
\clearpage

\begin{flushleft}
Table 1. Comparison of  energies  (in Ryd) and lifetimes (s) for the lowest 92 levels of  W LXVI. ($a{\pm}b \equiv$ $a\times$10$^{{\pm}b}$).
\end{flushleft}
\begin{tabular}{rllrrrrrrrrr} \hline
Index  & Configuration      & Level                   & GRASP3   & GRASP4     &$\tau$(GRASP1) & $\tau$(GRASP2) & $\tau$(GRASP3) &  $\tau$ (GRASP4)     \\
 \hline
    1  &    2s$^2$2p$^5$	    &   $^2$P$^o_{ 3/2}$      &   0.0000 &   0.0000   &   ........   &  ........    & ..........  & ......... \\
    2  &    2s$^2$2p$^5$	    &   $^2$P$^o_{ 1/2}$      & 102.0509 & 120.0486   &  4.138$-$11  &  4.137$-$11  & 4.137$-$11  & 4.17$-$11 \\
    3  &    2s2p$^6$	            &   $^2$S$  _{ 1/2}$      & 137.6754 & 137.6940   &  4.451$-$14  &  4.475$-$14  & 4.478$-$14  & 4.51$-$14 \\
    4  &    2s$^2$2p$^4$3s	    &   $^4$P$  _{ 5/2}$      & 621.0743 & 621.0697   &  2.596$-$14  &  2.609$-$14  & 2.611$-$14  & 2.61$-$14 \\
    5  &    2s$^2$2p$^4$3s	    &   $^2$P$  _{ 3/2}$      & 621.8872 & 621.8796   &  4.306$-$15  &  4.343$-$15  & 4.346$-$15  & 4.34$-$15 \\
    6  &    2s$^2$2p$^4$3s	    &   $^2$S$  _{ 1/2}$      & 625.8602 & 625.8603   &  8.686$-$15  &  8.753$-$15  & 8.761$-$15  & 8.76$-$15 \\
    7  &    2s$^2$2p$^4$($^3$P)3p   &   $^4$P$^o_{ 3/2}$      & 631.2412 & 631.2407   &  1.641$-$12  &  1.695$-$12  & 1.694$-$12  & 4.75$-$11 \\
    8  &    2s$^2$2p$^4$($^3$P)3p   &   $^2$D$^o_{ 5/2}$      & 631.3863 & 631.3844   &  5.693$-$13  &  5.792$-$13  & 5.796$-$13  & 7.44$-$11 \\
    9  &    2s$^2$2p$^4$($^1$S)3p   &   $^2$P$^o_{ 1/2}$      & 635.7706 & 635.7751   &  6.923$-$13  &  7.030$-$13  & 7.043$-$13  & 1.25$-$11 \\
   10  &    2s$^2$2p$^4$($^3$P)3p   &   $^4$P$^o_{ 5/2}$      & 659.8776 & 659.8754   &  4.762$-$13  &  4.781$-$13  & 4.785$-$13  & 1.19$-$12 \\
   11  &    2s$^2$2p$^4$($^3$P)3p   &   $^2$S$^o_{ 1/2}$      & 659.9483 & 659.9464   &  4.781$-$13  &  4.886$-$13  & 4.888$-$13  & 1.20$-$11 \\
   12  &    2s$^2$2p$^4$($^3$P)3p   &   $^4$D$^o_{ 7/2}$      & 659.8778 & 659.8753   &  4.742$-$13  &  4.759$-$13  & 4.763$-$13  & 6.13$-$13 \\
   13  &    2s$^2$2p$^4$($^3$P)3p   &   $^2$P$^o_{ 3/2}$      & 663.5191 & 663.5196   &  2.564$-$13  &  2.631$-$13  & 2.633$-$13  & 6.38$-$12 \\
   14  &    2s$^2$2p$^4$($^1$S)3p   &   $^2$P$^o_{ 3/2}$      & 665.4119 & 665.4188   &  5.199$-$13  &  5.170$-$13  & 5.173$-$13  & 2.63$-$09 \\
   15  &    2s$^2$2p$^4$($^1$D)3d   &   $^2$P$  _{ 3/2}$      & 670.7106 & 670.7140   &  4.125$-$14  &  4.133$-$14  & 4.132$-$14  & 4.36$-$14 \\
   16  &    2s$^2$2p$^4$($^3$P)3d   &   $^4$D$  _{ 5/2}$      & 670.8819 & 670.8850   &  1.320$-$14  &  1.330$-$14  & 1.332$-$14  & 1.35$-$14 \\
   17  &    2s$^2$2p$^4$($^3$P)3d   &   $^4$P$  _{ 1/2}$      & 671.1548 & 671.1567   &  2.857$-$15  &  2.862$-$15  & 2.863$-$15  & 2.87$-$15 \\
   18  &    2s$^2$2p$^4$($^3$P)3d   &   $^2$F$  _{ 7/2}$      & 670.9368 & 670.9401   &  8.388$-$13  &  8.375$-$13  & 8.385$-$13  & 8.41$-$13 \\
   19  &    2s$^2$2p$^4$($^1$S)3d   &   $^2$D$  _{ 3/2}$      & 675.3983 & 675.4067   &  1.566$-$13  &  1.563$-$13  & 1.555$-$13  & 1.94$-$13 \\
   20  &    2s$^2$2p$^4$($^3$P)3d   &   $^4$D$  _{ 7/2}$      & 677.3681 & 677.3713   &  4.171$-$12  &  4.173$-$12  & 4.177$-$12  & 9.36$-$10 \\
   21  &    2s$^2$2p$^4$($^3$P)3d   &   $^4$F$  _{ 9/2}$      & 677.4136 & 677.4175   &  7.795$-$12  &  7.811$-$12  & 7.827$-$12  & 7.82$-$12 \\
   22  &    2s$^2$2p$^4$($^3$P)3d   &   $^2$P$  _{ 1/2}$      & 678.4637 & 678.4638   &  1.267$-$15  &  1.272$-$15  & 1.271$-$15  & 1.27$-$15 \\
   23  &    2s$^2$2p$^4$($^3$P)3d   &   $^2$D$  _{ 5/2}$      & 680.2619 & 680.2594   &  6.140$-$16  &  6.162$-$16  & 6.173$-$16  & 6.16$-$16 \\
   24  &    2s$^2$2p$^4$($^3$P)3d   &   $^2$P$  _{ 3/2}$      & 680.3604 & 680.3573   &  4.950$-$16  &  5.004$-$16  & 5.005$-$16  & 5.01$-$16 \\
   25  &    2s$^2$2p$^4$($^1$S)3d   &   $^2$D$  _{ 5/2}$      & 682.9582 & 682.9646   &  8.968$-$16  &  9.163$-$16  & 9.142$-$16  & 9.17$-$16 \\
   26  &    2s$^2$2p$^4$3s	    &   $^4$P$  _{ 3/2}$      & 723.6503 & 723.6366   &  4.500$-$14  &  4.551$-$14  & 4.561$-$14  & 4.97$-$14 \\
   27  &    2s$^2$2p$^4$3s	    &   $^2$P$  _{ 1/2}$      & 724.3699 & 724.3534   &  4.774$-$15  &  4.812$-$15  & 4.815$-$15  & 1.96$-$14 \\
   28  &    2s$^2$2p$^4$3s	    &   $^2$D$  _{ 5/2}$      & 725.1435 & 725.1301   &  2.403$-$14  &  2.412$-$14  & 2.415$-$14  & 2.41$-$14 \\
   29  &    2s$^2$2p$^4$3s	    &   $^2$D$  _{ 3/2}$      & 725.6072 & 725.5919   &  5.647$-$15  &  5.685$-$15  & 5.688$-$15  & 5.06$-$14 \\
   30  &    2s$^2$2p$^4$($^3$P)3p   &   $^4$P$^o_{ 1/2}$      & 733.4481 & 733.4395   &  1.025$-$11  &  1.006$-$11  & 1.006$-$11  & 2.17$-$10 \\
   31  &    2s$^2$2p$^4$($^3$P)3p   &   $^4$D$^o_{ 3/2}$      & 733.9249 & 733.9161   &  1.447$-$12  &  1.493$-$12  & 1.492$-$12  & 1.31$-$11 \\
   32  &    2s$^2$2p$^4$($^1$D)3p   &   $^2$F$^o_{ 5/2}$      & 734.9602 & 734.9512   &  9.538$-$13  &  9.643$-$13  & 9.649$-$13  & 1.51$-$10 \\
   33  &    2s$^2$2p$^4$($^1$D)3p   &   $^2$P$^o_{ 3/2}$      & 738.0971 & 738.0933   &  6.761$-$13  &  7.436$-$13  & 7.427$-$13  & 1.19$-$12 \\
   34  &    2s2p$^5$($^3$P)3s	    &   $^4$P$^o_{ 5/2}$      & 754.0331 &            &  9.574$-$14  &  9.485$-$14  & 9.463$-$14  &           \\
   35  &    2s2p$^5$($^3$P)3s	    &   $^2$P$^o_{ 3/2}$      & 756.1273 &            &  6.259$-$15  &  6.303$-$15  & 6.329$-$15  &           \\
   36  &    2s2p$^5$($^1$P)3s	    &   $^2$P$^o_{ 1/2}$      & 760.3957 &            &  8.121$-$15  &  8.225$-$15  & 8.197$-$15  &           \\
   37  &    2s2p$^5$($^1$P)3s	    &   $^2$P$^o_{ 3/2}$      & 761.3648 & 	      &  5.081$-$14  &  5.101$-$14  & 5.080$-$14  &	      \\
   38  &    2s$^2$2p$^4$($^3$P)3p   &   $^4$D$^o_{ 5/2}$      & 762.5396 & 762.5287   &  3.750$-$13  &  3.772$-$13  & 3.774$-$13  & 2.80$-$12 \\
   39  &    2s$^2$2p$^4$($^3$P)3p   &   $^2$D$^o_{ 3/2}$      & 762.8995 & 762.8885   &  2.467$-$13  &  2.482$-$13  & 2.515$-$13  & 7.46$-$10 \\
   40  &    2s$^2$2p$^4$($^3$P)3p   &   $^2$P$^o_{ 1/2}$      & 763.3791 & 763.3714   &  1.376$-$14  &  1.369$-$14  & 1.381$-$14  & 1.61$-$14 \\
   41  &    2s$^2$2p$^4$($^1$D)3p   &   $^2$F$^o_{ 7/2}$      & 763.6864 & 763.6761   &  3.517$-$13  &  3.556$-$13  & 3.559$-$13  & 1.47$-$09 \\
   42  &    2s2p$^5$($^3$P)3p	    &   $^4$S$  _{ 3/2}$      & 764.7496 & 	      &  4.425$-$15  &  4.445$-$15  & 4.458$-$15  &	      \\
 \hline  											      
\end{tabular}   									   					       

\newpage
\begin{tabular}{rllrrrrrrrrrrr} \hline
Index  & Configuration      & Level                   & GRASP3   & GRASP4     &$\tau$(GRASP1) & $\tau$(GRASP2) & $\tau$(GRASP3) &  $\tau$ (GRASP4)     \\      
\hline      
   43  &    2s$^2$2p$^4$($^1$D)3p   &   $^2$D$^o_{ 3/2}$      & 764.6368 & 764.6281   &  8.337$-$14  &  8.170$-$14  & 8.194$-$14  & 2.49$-$13 \\
   44  &    2s2p$^5$($^3$P)3p	    &   $^2$D$  _{ 5/2}$      & 764.8120 & 764.6376   &  1.651$-$15  &  1.664$-$15  & 1.668$-$15  & 1.24$-$11 \\
   45  &    2s$^2$2p$^4$($^1$D)3p   &   $^2$D$^o_{ 5/2}$      & 764.6501 & 	      &  2.801$-$13  &  2.840$-$13  & 2.842$-$13  &	      \\
   46  &    2s$^2$2p$^4$($^1$D)3p   &   $^2$P$^o_{ 1/2}$      & 768.7551 & 768.7536   &  1.632$-$13  &  1.552$-$13  & 1.563$-$13  & 4.17$-$13 \\
   47  &    2s2p$^5$($^1$P)3p	    &   $^2$P$  _{ 1/2}$      & 770.8143 & 	      &  1.056$-$15  &  1.057$-$15  & 1.061$-$15  &	      \\
   48  &    2s2p$^5$($^1$P)3p	    &   $^2$D$  _{ 3/2}$      & 771.0121 & 	      &  1.792$-$15  &  1.793$-$15  & 1.800$-$15  &	      \\ 
   49  &    2s$^2$2p$^4$($^3$P)3d   &   $^4$D$  _{ 1/2}$      & 772.7321 & 772.7272   &  2.294$-$14  &  2.298$-$14  & 2.305$-$14  & 6.08$-$13 \\
   50  &    2s$^2$2p$^4$($^3$P)3d   &   $^4$D$  _{ 3/2}$      & 773.6008 & 773.5953   &  1.544$-$14  &  1.553$-$14  & 1.552$-$14  & 2.85$-$13 \\
   51  &    2s$^2$2p$^4$($^3$P)3d   &   $^4$F$  _{ 5/2}$      & 774.0316 & 774.0254   &  2.967$-$15  &  2.985$-$15  & 2.991$-$15  & 3.00$-$15 \\
   52  &    2s$^2$2p$^4$($^1$D)3d   &   $^2$G$  _{ 7/2}$      & 774.5477 & 774.5434   &  8.267$-$13  &  8.275$-$13  & 8.284$-$13  & 1.54$-$09 \\
   53  &    2s$^2$2p$^4$($^1$D)3d   &   $^2$S$  _{ 1/2}$      & 776.1650 & 776.1573   &  7.493$-$16  &  7.562$-$16  & 7.544$-$16  & 7.80$-$16 \\
   54  &    2s$^2$2p$^4$($^1$D)3d   &   $^2$F$  _{ 5/2}$      & 776.2201 & 776.2118   &  7.554$-$16  &  7.752$-$16  & 7.740$-$16  & 7.76$-$16 \\
   55  &    2s$^2$2p$^4$($^1$D)3d   &   $^2$D$  _{ 3/2}$      & 776.6317 & 776.6228   &  6.662$-$16  &  6.799$-$16  & 6.788$-$16  & 6.89$-$16 \\
   56  &    2s$^2$2p$^4$($^3$P)3d   &   $^4$F$  _{ 7/2}$      & 779.7930 & 779.7869   &  4.364$-$12  &  4.317$-$12  & 4.321$-$12  & 4.09$-$07 \\
   57  &    2s$^2$2p$^4$($^3$P)3d   &   $^2$F$  _{ 5/2}$      & 780.8165 & 780.8088   &  3.207$-$12  &  3.216$-$12  & 3.221$-$12  & 1.85$-$11 \\
   58  &    2s$^2$2p$^4$($^3$P)3d   &   $^4$P$  _{ 3/2}$      & 780.8796 & 780.8718   &  9.657$-$15  &  9.552$-$15  & 9.590$-$15  & 6.67$-$12 \\
   59  &    2s$^2$2p$^4$($^1$D)3d   &   $^2$G$  _{ 9/2}$      & 781.2936 & 781.2890   &  7.065$-$12  &  7.088$-$12  & 7.100$-$12  & 3.82$-$09 \\
   60  &    2s$^2$2p$^4$($^1$D)3d   &   $^2$D$  _{ 5/2}$      & 781.9634 & 781.9558   &  2.757$-$12  &  2.917$-$12  & 2.915$-$12  & 2.64$-$11 \\
   61  &    2s$^2$2p$^4$($^1$D)3d   &   $^2$F$  _{ 7/2}$      & 782.4471 & 782.4398   &  5.584$-$12  &  5.496$-$12  & 5.504$-$12  & 9.81$-$10 \\
   62  &    2s$^2$2p$^4$($^3$P)3d   &   $^2$D$  _{ 3/2}$      & 784.1419 & 784.1285   &  3.466$-$16  &  3.510$-$16  & 3.509$-$16  & 5.98$-$14 \\
   63  &    2s$^2$2p$^4$($^1$D)3d   &   $^2$P$  _{ 1/2}$      & 784.6003 & 784.5867   &  3.221$-$16  &  3.254$-$16  & 3.253$-$16  & 3.55$-$14 \\
   64  &    2s2p$^5$($^3$P)3p	    &   $^4$D$  _{ 7/2}$      & 793.1551 & 	      &  8.336$-$14  &  8.385$-$14  & 8.375$-$14  &	      \\
   65  &    2s2p$^5$($^3$P)3p	    &   $^2$P$  _{ 3/2}$      & 793.7276 & 	      &  2.920$-$15  &  2.942$-$15  & 2.957$-$15  &	      \\
   66  &    2s2p$^5$($^3$P)3p	    &   $^4$P$  _{ 5/2}$      & 794.2004 & 	      &  4.455$-$15  &  4.501$-$15  & 4.513$-$15  &	      \\
   67  &    2s2p$^5$($^3$P)3p	    &   $^2$P$  _{ 1/2}$      & 795.8038 & 	      &  2.122$-$15  &  2.141$-$15  & 2.149$-$15  &	      \\
   68  &    2s2p$^5$($^1$P)3p	    &   $^2$D$  _{ 5/2}$      & 799.7800 & 	      &  3.539$-$15  &  3.540$-$15  & 3.548$-$15  &	      \\
   69  &    2s2p$^5$($^1$P)3p	    &   $^2$P$  _{ 3/2}$      & 800.2864 & 	      &  5.631$-$15  &  5.686$-$15  & 5.681$-$15  &	      \\
   70  &    2s2p$^5$($^3$P)3p	    &   $^2$S$  _{ 1/2}$      & 802.5328 & 	      &  2.209$-$14  &  2.204$-$14  & 2.202$-$14  &	      \\
   71  &    2s2p$^5$($^3$P)3d	    &   $^4$P$^o_{ 1/2}$      & 803.4929 & 	      &  2.775$-$14  &  2.781$-$14  & 2.788$-$14  &	      \\
   72  &    2s2p$^5$($^3$P)3d	    &   $^4$P$^o_{ 3/2}$      & 804.1891 & 	      &  1.209$-$14  &  1.214$-$14  & 1.218$-$14  &	      \\
   73  &    2s2p$^5$($^3$P)3d	    &   $^2$F$^o_{ 7/2}$      & 804.3974 & 	      &  5.838$-$14  &  6.013$-$14  & 6.013$-$14  &	      \\
   74  &    2s2p$^5$($^3$P)3d	    &   $^4$D$^o_{ 5/2}$      & 804.7064 & 	      &  7.001$-$14  &  7.280$-$14  & 7.283$-$14  &	      \\
   75  &    2s2p$^5$($^1$P)3d	    &   $^2$P$^o_{ 1/2}$      & 810.4719 & 	      &  1.039$-$14  &  1.021$-$14  & 1.020$-$14  &	      \\
   76  &    2s2p$^5$($^1$P)3d	    &   $^2$F$^o_{ 5/2}$      & 810.3798 & 	      &  3.879$-$14  &  4.024$-$14  & 4.016$-$14  &	      \\
   77  &    2s2p$^5$($^1$P)3d	    &   $^2$D$^o_{ 3/2}$      & 810.6739 & 	      &  2.017$-$14  &  2.074$-$14  & 2.078$-$14  &	      \\
   78  &    2s2p$^5$($^3$P)3d	    &   $^4$F$^o_{ 9/2}$      & 810.2333 & 	      &  9.542$-$14  &  1.004$-$13  & 1.003$-$13  &	      \\
   79  &    2s2p$^5$($^3$P)3d	    &   $^4$D$^o_{ 7/2}$      & 811.7811 & 	      &  6.573$-$14  &  6.771$-$14  & 6.774$-$14  &	      \\
   80  &    2s2p$^5$($^3$P)3d	    &   $^2$D$^o_{ 5/2}$      & 811.8840 & 	      &  4.485$-$14  &  4.551$-$14  & 4.566$-$14  &	      \\
   81  &    2s2p$^5$($^3$P)3d	    &   $^2$P$^o_{ 3/2}$      & 813.3103 & 	      &  1.241$-$15  &  1.253$-$15  & 1.258$-$15  &	      \\
   82  &    2s2p$^5$($^3$P)3d	    &   $^2$P$^o_{ 1/2}$      & 814.7743 & 	      &  3.470$-$16  &  3.514$-$16  & 3.526$-$16  &	      \\
   83  &    2s2p$^5$($^1$P)3d	    &   $^2$F$^o_{ 7/2}$      & 817.4226 & 	      &  2.963$-$14  &  3.040$-$14  & 3.039$-$14  &	      \\
   84  &    2s2p$^5$($^1$P)3d	    &   $^2$D$^o_{ 5/2}$      & 818.0069 & 	      &  3.210$-$14  &  3.311$-$14  & 3.309$-$14  &	      \\
   85  &    2s2p$^5$($^1$P)3d	    &   $^2$P$^o_{ 3/2}$      & 819.0259 & 	      &  4.533$-$16  &  4.587$-$16  & 4.597$-$16  &	      \\
 \hline  											      
\end{tabular}   									   					       

\newpage
\begin{tabular}{rllrrrrrrrrrrr} \hline
Index  & Configuration      & Level                   & GRASP3   & GRASP4     &$\tau$(GRASP1) & $\tau$(GRASP2) & $\tau$(GRASP3) &  $\tau$ (GRASP4)     \\   
\hline   
   86  &    2s$^2$2p$^4$3s	    &   $^4$P$  _{ 1/2}$      & 829.7354 & 829.7138   &  1.955$-$14  &  1.966$-$14  & 1.969$-$14  & 8.87$-$11 \\
   87  &    2s$^2$2p$^4$($^3$P)3p   &   $^4$D$^o_{ 1/2}$      & 840.8018 & 840.7876   &  1.340$-$12  &  1.365$-$12  & 1.359$-$12  & 2.28$-$12 \\
   88  &    2s2p$^5$($^3$P)3s	    &   $^4$P$^o_{ 1/2}$      & 858.3681 & 	      &  2.269$-$14  &  2.263$-$14  & 2.264$-$14  &	      \\
   89  &    2s2p$^5$($^3$P)3s	    &   $^4$P$^o_{ 3/2}$      & 860.6983 & 	      &  1.575$-$14  &  1.574$-$14  & 1.573$-$14  &	      \\
   90  &    2s2p$^5$($^3$P)3s	    &   $^2$P$^o_{ 1/2}$      & 862.9096 & 	      &  3.035$-$14  &  3.075$-$14  & 3.093$-$14  &	      \\
   91  &    2s2p$^5$($^3$P)3p	    &   $^4$D$  _{ 1/2}$      & 868.3439 & 	      &  5.079$-$15  &  5.095$-$15  & 5.099$-$15  &	      \\
   92  &    2s$^2$2p$^4$($^3$P)3p   &   $^4$S$^o_{ 3/2}$      & 868.6480 & 868.6295   &  1.009$-$13  &  1.022$-$13  & 1.025$-$13  & 1.78$-$13 \\
\hline				  				        				   
\end{tabular}										        		   
\begin {flushleft}									        		   
\begin{tabbing} 									      
aaaaaaaaaaaaaaaaaaaaaaaaaaaaaaaaaaaa\= \kill						      		      
GRASP1: Present results with the {\sc grasp} code from 11  configurations and 113 levels \\
GRASP2: Present results with the {\sc grasp} code from 23  configurations and 359 levels \\
GRASP3: Present results with the {\sc grasp} code from 38  configurations and 501 levels \\
GRASP4: earlier results of S.~Aggarwal$^{\cite{sa}}$ with the {\sc grasp} code from 23  configurations and 359 levels \\
		      		      
\end{tabbing}										      
\end {flushleft}				      							   					       


\newpage
\clearpage
\begin{flushleft}
Table 2.  Total and dominant A-values (s$^{-1}$) for the lowest 92 levels of  W LXVI. ($a{\pm}b \equiv$ $a\times$10$^{{\pm}b}$).
\end{flushleft}
{\footnotesize 
\begin{tabular}{rllllll} \hline
Index  &     Configuration          & Level                   &   ${\sum_{i} A_{ji}}$ & Dominant A$_{ji}$   \\
\hline
    1  &    2s$^2$2p$^5$	    &   $^2$P$^o_{ 3/2}$      &    ...........  \\
    2  &    2s$^2$2p$^5$	    &   $^2$P$^o_{ 1/2}$      &    2.417$+$10 &  1 --	 2   M1:   2.396$+$10									 \\
    3  &    2s2p$^6$	            &   $^2$S$  _{ 1/2}$      &    2.233$+$13 &  1 --	 3   E1:   2.216$+$13									 \\
    4  &    2s$^2$2p$^4$3s	    &   $^4$P$  _{ 5/2}$      &    3.831$+$13 &  1 --	 4   E1:   3.830$+$13									 \\
    5  &    2s$^2$2p$^4$3s	    &   $^2$P$  _{ 3/2}$      &    2.301$+$14 &  1 --	 5   E1:   2.301$+$14									 \\
    6  &    2s$^2$2p$^4$3s	    &   $^2$S$  _{ 1/2}$      &    1.141$+$14 &  1 --	 6   E1:   1.141$+$14									 \\
    7  &    2s$^2$2p$^4$($^3$P)3p   &   $^4$P$^o_{ 3/2}$      &    5.902$+$11 &  1 --	 7   E2:   4.955$+$11									 \\
    8  &    2s$^2$2p$^4$($^3$P)3p   &   $^2$D$^o_{ 5/2}$      &    1.725$+$12 &  1 --	 8   E2:   1.692$+$12									 \\
    9  &    2s$^2$2p$^4$($^1$S)3p   &   $^2$P$^o_{ 1/2}$      &    1.420$+$12 &  1 --	 9   E2:   1.291$+$12									 \\
   10  &    2s$^2$2p$^4$($^3$P)3p   &   $^4$P$^o_{ 5/2}$      &    2.090$+$12 &  1 --	10   E2:   5.174$+$11, 4 --  10     E1:   8.411$+$11, 5 --  10     E1:   7.228$+$11	 \\
   11  &    2s$^2$2p$^4$($^3$P)3p   &   $^2$S$^o_{ 1/2}$      &    2.046$+$12 &  1 --	11   E2:   4.393$+$11, 5 --  11     E1:   1.503$+$12					 \\
   12  &    2s$^2$2p$^4$($^3$P)3p   &   $^4$D$^o_{ 7/2}$      &    2.100$+$12 &  1 --	12   E2:   4.686$+$11, 4 --  12     E1:   1.631$+$12					 \\
   13  &    2s$^2$2p$^4$($^3$P)3p   &   $^2$P$^o_{ 3/2}$      &    3.799$+$12 &  1 --	13   E2:   1.626$+$12, 5 --  13     E1:   1.419$+$12					 \\
   14  &    2s$^2$2p$^4$($^1$S)3p   &   $^2$P$^o_{ 3/2}$      &    1.933$+$12 &  6 --	14   E1:   1.476$+$12									 \\
   15  &    2s$^2$2p$^4$($^1$D)3d   &   $^2$P$  _{ 3/2}$      &    2.420$+$13 &  1 --	15   E1:   2.295$+$13									 \\
   16  &    2s$^2$2p$^4$($^3$P)3d   &   $^4$D$  _{ 5/2}$      &    7.506$+$13 &  1 --	16   E1:   7.388$+$13									 \\
   17  &    2s$^2$2p$^4$($^3$P)3d   &   $^4$P$  _{ 1/2}$      &    3.493$+$14 &  1 --	17   E1:   3.480$+$14									 \\
   18  &    2s$^2$2p$^4$($^3$P)3d   &   $^2$F$  _{ 7/2}$      &    1.193$+$12 &  8 --	18   E1:   1.187$+$12									 \\
   19  &    2s$^2$2p$^4$($^1$S)3d   &   $^2$D$  _{ 3/2}$      &    6.429$+$12 &  1 --	19   E1:   5.193$+$12									 \\
   20  &    2s$^2$2p$^4$($^3$P)3d   &   $^4$D$  _{ 7/2}$      &    2.394$+$11 &  1 --	20   M2:   1.118$+$11,10 --  20     E1:   9.891$+$10					 \\
   21  &    2s$^2$2p$^4$($^3$P)3d   &   $^4$F$  _{ 9/2}$      &    1.278$+$11 & 12 --	21   E1:   1.276$+$11									 \\
   22  &    2s$^2$2p$^4$($^3$P)3d   &   $^2$P$  _{ 1/2}$      &    7.865$+$14 &  1 --	22   E1:   7.862$+$14									 \\
   23  &    2s$^2$2p$^4$($^3$P)3d   &   $^2$D$  _{ 5/2}$      &    1.620$+$15 &  1 --	23   E1:   1.620$+$15									 \\
   24  &    2s$^2$2p$^4$($^3$P)3d   &   $^2$P$  _{ 3/2}$      &    1.998$+$15 &  1 --	24   E1:   1.998$+$15									 \\
   25  &    2s$^2$2p$^4$($^1$S)3d   &   $^2$D$  _{ 5/2}$      &    1.094$+$15 &  1 --	25   E1:   1.094$+$15									 \\
   26  &    2s$^2$2p$^4$3s	    &   $^4$P$  _{ 3/2}$      &    2.192$+$13 &  1 --	26   E1:   2.007$+$13									 \\
   27  &    2s$^2$2p$^4$3s	    &   $^2$P$  _{ 1/2}$      &    2.077$+$14 &  1 --	27   E1:   5.106$+$13, 2 --  27     E1:   1.566$+$14					 \\
   28  &    2s$^2$2p$^4$3s	    &   $^2$D$  _{ 5/2}$      &    4.141$+$13 &  1 --	28   E1:   4.139$+$13									 \\
   29  &    2s$^2$2p$^4$3s	    &   $^2$D$  _{ 3/2}$      &    1.758$+$14 &  2 --	29   E1:   1.560$+$14									 \\
   30  &    2s$^2$2p$^4$($^3$P)3p   &   $^4$P$^o_{ 1/2}$      &    9.945$+$10 &  2 --	30   M1:   3.243$+$10, 7 --  30     M1:   2.043$+$10, 26 --  30     E1:   2.074$+$10	 \\
   31  &    2s$^2$2p$^4$($^3$P)3p   &   $^4$D$^o_{ 3/2}$      &    6.703$+$11 &  2 --	31   E2:   5.151$+$11									 \\
   32  &    2s$^2$2p$^4$($^1$D)3p   &   $^2$F$^o_{ 5/2}$      &    1.036$+$12 &  2 --	32   E2:   9.909$+$11									 \\
   33  &    2s$^2$2p$^4$($^1$D)3p   &   $^2$P$^o_{ 3/2}$      &    1.346$+$12 &  2 --	33   E2:   4.710$+$11, 3 --  33     E1:   6.239$+$11					 \\
   34  &    2s2p$^5$($^3$P)3s	    &   $^4$P$^o_{ 5/2}$      &    1.057$+$13 &  4 --	34   E1:   9.820$+$12									 \\
   35  &    2s2p$^5$($^3$P)3s	    &   $^2$P$^o_{ 3/2}$      &    1.580$+$14 &  3 --	35   E1:   1.467$+$14									 \\
   36  &    2s2p$^5$($^1$P)3s	    &   $^2$P$^o_{ 1/2}$      &    1.220$+$14 &  3 --	36   E1:   1.044$+$14									 \\
   37  &    2s2p$^5$($^1$P)3s	    &   $^2$P$^o_{ 3/2}$      &    1.968$+$13 &  4 --	37   E1:   1.012$+$13,  6 --   37   E1:   4.484$+$12					 \\
   38  &    2s$^2$2p$^4$($^3$P)3p   &   $^4$D$^o_{ 5/2}$      &    2.649$+$12 & 26 --	38   E1:   1.711$+$12									 \\
   39  &    2s$^2$2p$^4$($^3$P)3p   &   $^2$D$^o_{ 3/2}$      &    3.976$+$12 &  1 --	39   E2:   1.102$+$12									 \\
   40  &    2s$^2$2p$^4$($^3$P)3p   &   $^2$P$^o_{ 1/2}$      &    7.241$+$13 &  3 --	40   E1:   6.132$+$13									 \\
   41  &    2s$^2$2p$^4$($^1$D)3p   &   $^2$F$^o_{ 7/2}$      &    2.810$+$12 &  1 --	41   E2:   1.205$+$12,  28 -- 41    E1:   1.590$+$12					 \\
   42  &    2s2p$^5$($^3$P)3p	    &   $^4$S$  _{ 3/2}$      &    2.243$+$14 &  1 --	42   E1:   2.131$+$14									 \\
   43  &    2s$^2$2p$^4$($^1$D)3p   &   $^2$D$^o_{ 3/2}$      &    1.220$+$13 &  3 --	43   E1:   4.034$+$12,  4 -- 43     E1:   4.041$+$12					 \\
   44  &    2s2p$^5$($^3$P)3p	    &   $^2$D$  _{ 5/2}$      &    5.996$+$14 &  1 --	44   E1:   5.881$+$14	     \\ 							 
   45  &    2s$^2$2p$^4$($^1$D)3p   &   $^2$D$^o_{ 5/2}$      &    3.519$+$12 &  1 --	45   E2:   7.593$+$11,  2 -- 45     E2:   8.964$+$11, 28 --  45     E1:   9.669$+$11, 29  -- 45     E1:   7.800$+$11  \\
   46  &    2s$^2$2p$^4$($^1$D)3p   &   $^2$P$^o_{ 1/2}$      &    6.397$+$12 &  3 --	46   E1:   2.358$+$12, 29 -- 46     E1:   2.004$+$12								      \\
   47  &    2s2p$^5$($^1$P)3p	    &   $^2$P$  _{ 1/2}$      &    9.426$+$14 &  1 --	47   E1:   9.142$+$14												      \\
   48  &    2s2p$^5$($^1$P)3p	    &   $^2$D$  _{ 3/2}$      &    5.556$+$14 &  1 --	48   E1:   5.289$+$14												      \\
\hline	
\end{tabular}
}
\newpage
{\footnotesize 
\begin{tabular}{rllllll} \hline
Index  &     Configuration                        & Level    &   ${\sum_{i} A_{ji}}$ & Dominant A$_{ji}$   \\  
\hline    
   49  &    2s$^2$2p$^4$($^3$P)3d   &   $^4$D$  _{ 1/2}$      &    4.339$+$13 &  2 --	49   E1:   4.062$+$13												      \\
   50  &    2s$^2$2p$^4$($^3$P)3d   &   $^4$D$  _{ 3/2}$      &    6.442$+$13 &  2 --	50   E1:   5.933$+$13												      \\
   51  &    2s$^2$2p$^4$($^3$P)3d   &   $^4$F$  _{ 5/2}$      &    3.344$+$14 &  1 --	51   E1:   3.330$+$14												      \\
   52  &    2s$^2$2p$^4$($^1$D)3d   &   $^2$G$  _{ 7/2}$      &    1.207$+$12 & 32 --	52   E1:   1.185$+$12												      \\
   53  &    2s$^2$2p$^4$($^1$D)3d   &   $^2$S$  _{ 1/2}$      &    1.326$+$15 &  1 --	53   E1:   1.286$+$15												      \\
   54  &    2s$^2$2p$^4$($^1$D)3d   &   $^2$F$  _{ 5/2}$      &    1.292$+$15 &  1 --	54   E1:   1.290$+$15												      \\
   55  &    2s$^2$2p$^4$($^1$D)3d   &   $^2$D$  _{ 3/2}$      &    1.473$+$15 &  1 --	55   E1:   1.454$+$15												      \\
   56  &    2s$^2$2p$^4$($^3$P)3d   &   $^4$F$  _{ 7/2}$      &    2.314$+$11 & 38 --	56   E1:   1.074$+$11												      \\
   57  &    2s$^2$2p$^4$($^3$P)3d   &   $^2$F$  _{ 5/2}$      &    3.105$+$11 &  2 --	57   M2:   6.951$+$10, 39 --57     E1:   9.236$+$10								      \\
   58  &    2s$^2$2p$^4$($^3$P)3d   &   $^4$P$  _{ 3/2}$      &    1.043$+$14 &  2 --	58   E1:   1.038$+$14												      \\
   59  &    2s$^2$2p$^4$($^1$D)3d   &   $^2$G$  _{ 9/2}$      &    1.408$+$11 & 41 --	59   E1:   1.278$+$11												      \\
   60  &    2s$^2$2p$^4$($^1$D)3d   &   $^2$D$  _{ 5/2}$      &    3.430$+$11 &  1 --	60   E1:   3.845$+$10, 2 -- 60     M2:   6.058$+$10, 43  -- 60     E1:   4.924$+$10, 45 --  60     E1:   4.317$+$10   \\
   61  &    2s$^2$2p$^4$($^1$D)3d   &   $^2$F$  _{ 7/2}$      &    1.817$+$11 & 45 --	61   E1:   9.946$+$10												      \\
   62  &    2s$^2$2p$^4$($^3$P)3d   &   $^2$D$  _{ 3/2}$      &    2.850$+$15 &  2 --	62   E1:   2.833$+$15												      \\
   63  &    2s$^2$2p$^4$($^1$D)3d   &   $^2$P$  _{ 1/2}$      &    3.074$+$15 &  2 --	63   E1:   3.045$+$15												      \\
   64  &    2s2p$^5$($^3$P)3p	    &   $^4$D$  _{ 7/2}$      &    1.194$+$13 & 12 --	64   E1:   8.936$+$12												      \\
   65  &    2s2p$^5$($^3$P)3p	    &   $^2$P$  _{ 3/2}$      &    3.382$+$14 &  1 --	65   E1:   3.239$+$14												      \\
   66  &    2s2p$^5$($^3$P)3p	    &   $^4$P$  _{ 5/2}$      &    2.216$+$14 &  1 --	66   E1:   2.081$+$14												      \\
   67  &    2s2p$^5$($^3$P)3p	    &   $^2$P$  _{ 1/2}$      &    4.654$+$14 &  1 --	67   E1:   4.502$+$14												      \\
   68  &    2s2p$^5$($^1$P)3p	    &   $^2$D$  _{ 5/2}$      &    2.819$+$14 &  1 --	68   E1:   2.536$+$14												      \\
   69  &    2s2p$^5$($^1$P)3p	    &   $^2$P$  _{ 3/2}$      &    1.760$+$14 &  1 --	69   E1:   1.455$+$14												      \\
   70  &    2s2p$^5$($^3$P)3p	    &   $^2$S$  _{ 1/2}$      &    4.541$+$13 &  1 --	70   E1:   1.216$+$13, 14 -- 70     E1:   1.380$+$13								      \\
   71  &    2s2p$^5$($^3$P)3d	    &   $^4$P$^o_{ 1/2}$      &    3.587$+$13 &  3 --	71   E1:   2.379$+$13												      \\
   72  &    2s2p$^5$($^3$P)3d	    &   $^4$P$^o_{ 3/2}$      &    8.209$+$13 &  3 --	72   E1:   6.951$+$13												      \\
   73  &    2s2p$^5$($^3$P)3d	    &   $^2$F$^o_{ 7/2}$      &    1.663$+$13 &  1 --	73   E2:   5.508$+$12, 18 -- 73     E1:   8.687$+$12								      \\
   74  &    2s2p$^5$($^3$P)3d	    &   $^4$D$^o_{ 5/2}$      &    1.373$+$13 & 16 --	74   E1:   6.176$+$12												      \\
   75  &    2s2p$^5$($^1$P)3d	    &   $^2$P$^o_{ 1/2}$      &    9.807$+$13 &  3 --	75   E1:   6.424$+$13												      \\
   76  &    2s2p$^5$($^1$P)3d	    &   $^2$F$^o_{ 5/2}$      &    2.490$+$13 & 18 --	76   E1:   1.437$+$13, 19 -- 76     E1:   5.022$+$12								      \\
   77  &    2s2p$^5$($^1$P)3d	    &   $^2$D$^o_{ 3/2}$      &    4.811$+$13 &  3 --	77   E1:   2.378$+$13, 16 -- 77     E1:   9.757$+$12								      \\
   78  &    2s2p$^5$($^3$P)3d	    &   $^4$F$^o_{ 9/2}$      &    9.967$+$12 & 21 --	78   E1:   8.177$+$12												      \\
   79  &    2s2p$^5$($^3$P)3d	    &   $^4$D$^o_{ 7/2}$      &    1.476$+$13 &  1 --	79   E2:   4.504$+$12, 20 -- 79     E1:   4.450$+$12								      \\
   80  &    2s2p$^5$($^3$P)3d	    &   $^2$D$^o_{ 5/2}$      &    2.190$+$13 &  1 --	80   E2:   8.730$+$12, 20 -- 80     E1:   6.393$+$12								      \\
   81  &    2s2p$^5$($^3$P)3d	    &   $^2$P$^o_{ 3/2}$      &    7.950$+$14 &  3 --	81   E1:   7.656$+$14												      \\
   82  &    2s2p$^5$($^3$P)3d	    &   $^2$P$^o_{ 1/2}$      &    2.836$+$15 &  3 --	82   E1:   2.816$+$15												      \\
   83  &    2s2p$^5$($^1$P)3d	    &   $^2$F$^o_{ 7/2}$      &    3.290$+$13 &  1 --	83   E2:   7.486$+$12, 21 -- 83     E1:   1.473$+$13								      \\
   84  &    2s2p$^5$($^1$P)3d	    &   $^2$D$^o_{ 5/2}$      &    3.022$+$13 & 20 --	84   E1:   7.795$+$12, 23 -- 84     E1:   9.119$+$12								      \\
   85  &    2s2p$^5$($^1$P)3d	    &   $^2$P$^o_{ 3/2}$      &    2.176$+$15 &  3 --	85   E1:   2.151$+$15												      \\
   86  &    2s$^2$2p$^4$3s	    &   $^4$P$  _{ 1/2}$      &    5.080$+$13 &  2 --	86   E1:   5.066$+$13												      \\
   87  &    2s$^2$2p$^4$($^3$P)3p   &   $^4$D$^o_{ 1/2}$      &    7.361$+$11 &  3 --	87   E1:   4.424$+$11												      \\
   88  &    2s2p$^5$($^3$P)3s	    &   $^4$P$^o_{ 1/2}$      &    4.417$+$13 &  3 --	88   E1:   2.222$+$13, 26 -- 88     E1:   1.676$+$13								      \\
   89  &    2s2p$^5$($^3$P)3s	    &   $^4$P$^o_{ 3/2}$      &    6.357$+$13 &  3 --	89   E1:   4.063$+$13, 28 -- 89     E1:   1.730$+$13								      \\
   90  &    2s2p$^5$($^3$P)3s	    &   $^2$P$^o_{ 1/2}$      &    3.233$+$13 &  3 --	90   E1:   1.051$+$13, 29 -- 90     E1:   1.674$+$13								      \\
   91  &    2s2p$^5$($^3$P)3p	    &   $^4$D$  _{ 1/2}$      &    1.961$+$14 &  2 --	91   E1:   1.745$+$14												      \\
   92  &    2s$^2$2p$^4$($^3$P)3p   &   $^4$S$^o_{ 3/2}$      &    9.760$+$12 &  3 --	92   E1:   5.596$+$12												      \\
\hline
\end{tabular}	
}									        				   
			       
\end{document}